\documentclass[twocolumn,epsfig,superscriptaddressshowpacs]{revtex4}
\usepackage{bm} 
\usepackage{hyperref}
\def\beq{\begin{equation}}
\def\eeq{\end{equation}}

\def\etal{{\it et al.}}

\def\beq{\begin{equation}}                           
\def\eeq{\end{equation}}                           
\def\bea{\begin{eqnarray}}                           
\def\eea{\end{eqnarray}}

\newcommand{\bsf}[1]{\textsf{\textbf{#1}}}
\draft
                   
\begin{document}
\textwidth = 6.7 in
\textheight = 9 in
\oddsidemargin = 0.0 in
\evensidemargin = 0.0 in
\topmargin = 0.0 in
\headheight = 0.0 in
\headsep = 0.0 in
\parskip = 0.2in
\parindent = 0.0in

\title{The Mechanics and Statistics of Active Matter}
\author{Sriram Ramaswamy}%
\email{sriram@physics.iisc.ernet.in}
\affiliation{Centre for Condensed Matter Theory, Department of Physics, Indian Institute of Science, 
Bangalore 560 012 India}
\altaffiliation{Also with CMTU, JNCASR, Bangalore 560 064}

\date{\today}

\begin{abstract}
{Active particles contain internal degrees of freedom with the ability to take
in and dissipate energy and, in the process, execute systematic movement.
Examples include all living organisms and their motile constituents such as
molecular motors. This article reviews recent progress in applying the
principles of nonequilibrium statistical mechanics and hydrodynamics to form a
systematic theory of the behaviour of collections of active particles -- active
matter -- with only minimal regard to microscopic details. A unified view of
the many kinds of active matter is presented, encompassing not only living
systems but inanimate analogues. Theory and experiment are discussed side by
side. {\bf This review is to appear in volume 1 of the Annual Review of Condensed Matter Physics in July 2010 and is posted here with permission from that journal}}
\end{abstract}

\maketitle

%
%
%
%
\input psfig.sty
%
%
%
%
%
%
%
%

\section{INTRODUCTION} 
\label{intro} 
Physics enters biology in two broad and overlapping areas -- information and
mechanics \cite{philnelsonbook,robphillipsbook}. This review is about the
mechanics, statistical and otherwise, of living matter. This is not a review of
soft-matter physics in a biological setting. The aspect of biological matter of
interest here is the ability to transduce free energy into systematic movement.
This property is the defining characteristic of active matter, and my interest
is in the unique mechanical properties that arise as a consequence of such
active processes \cite{alberts}. This review focusses on the collective
behaviour of systems with a large number of constituents, and will thus use the
ideas of condensed matter and statistical physics. I will not discuss the physics 
of molecular motors or
the related and vast exclusion-process family of problems \cite{kolomeisky_fisher,frey09,debchow}. 
The aim is to complement and
update the perspectives of earlier reviews \cite{tonertusr,physrep} and to
present the achievements and limitations of this rapidly advancing subfield.
I begin with the realizations of active matter to be covered.

\subsection{Systems of Interest: Flocks, Rods, and the Cytoskeleton}
\label{sysofint}
\subsubsection{Flocks: theoretical, real, and imitation}
\label{flocksub}
It is natural for a condensed matter physicist to regard a coherently moving
flock of birds, beasts or bacteria as an orientationally ordered phase of
living matter. This idea was first implemented \cite{reynolds,vicsek1} in
computer models of interacting particles moving with a fixed speed and trying
to align their velocity vectors parallel to those of their neighbours, in the
presence of noise. To a student of magnetism these are itinerant, classical,
ferromagnetically interacting continuous spins that move in the
direction in which they point, which is what makes flocks different
from magnets. The models showed a nonequilibrium phase transition from a
disordered state to a flock with long-range order \cite{vicsek1,tonertu,tonertuprl95,tutonerulm} 
in the particle velocities as the noise strength was decreased or the concentration of
particles raised. The nature of order and fluctuations in a flock, the
character of the transition to a flock, and the flocking of active particles
with an axis but no polarity are among the issues this review deals with. 

Amazingly, the physics of flocking can be imitated by a collection of rods
lying on a horizontal surface, agitated vertically \cite{sradititoner}. Indeed,
vibrated monolayers of macroscopic grains have provided some of the most
fruitful realizations of active matter. This review will explain briefly the
physics behind underlying this analogy and highlight experiments that exploit
it \cite{vjnmsrscience} 

Flocking in a fluid medium introduces physics absent in the simplest flocking
models: each swimming creature generates fluid flow which moves and reorients
other swimmers far away. Including this interaction leads \cite{aditisr} to a
modified liquid-crystal hydrodynamics \cite{degp}, in which the constituent
particles carry permanent stresses that stir the fluid. The implications of
this interplay of self-propelling activity and fluid flow for order and
macroscopic rheology are an important part of this review.  

\subsubsection{The cytoskeleton as an active gel}
\label{sub:cyto} 
\begin{figure}
\centerline{\psfig{figure=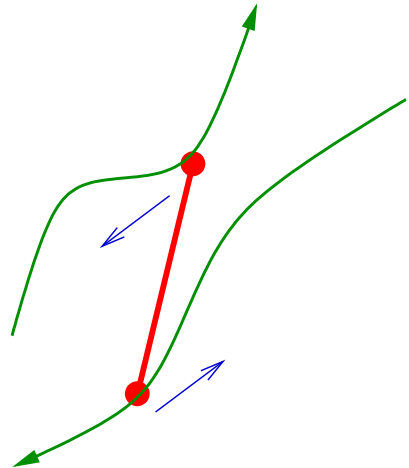,height=15pc}}
\caption{A cluster of motors with heads on both sides, exerting contractile forces on the cytoskeletal network.}
\label{contractfig}
\end{figure}

The physics of the flocking of organisms in a fluid reappears at a
\textit{sub}cellular scale in the cytoskeleton \cite{alberts,kaeswebsite}, the
polymeric scaffolding that governs transport, adhesion, movement and division
in the living cell.  Two nonequilibrium processes drive the cytoskeleton: the
ATP-assisted polymerization and depolymerization known as
\textit{treadmilling}, which I will not discuss; and \textit{contractility},
illustrated in Fig. \ref{contractfig}, which arises from the ATP-driven movement of
motor proteins in specific directions on biofilaments, and is central to this
review.  The cytoskeleton is thus a suspension of filaments endowed with active
internal forces. Moreover, there are natural mechanisms that promote the
alignment of neighbouring filaments, through excluded volume as well as
activity. We must therefore allow for the possibility of orientational order. 
It is then not surprising that the hydrodynamic equations obtained in the active-gel
description \cite{activegel1,activegel2} of the cytoskeleton have precisely the
same form as those \cite{aditisr} for collections of swimming organisms,
ignoring complications such as permanent crosslinking. 
A condensed-matter physicist accustomed to symmetry arguments should find it reassuring that identical hydrodynamic descriptions apply for a $10$ km fish shoal and a $10$
$\mu$m cytoskeleton. 

\subsection{Viewpoint: active systems as material}
\label{activematerialsub}
The viewpoint of this review is that living matter can fruitfully be regarded
as a kind of material \cite{cellmaterial} and studied using the tools of
condensed matter physics and statistical mechanics; that there is a practical
way to encode into such a description those features of the living state that
are relevant to materials science; and that the results of such an endeavour
will help us better understand, control and perhaps mimic active cellular
matter.  A community of respectable size has grown around this activity,
approaching the problem at different length scales and with a variety of
techniques. What do they, and we, hope to gain from this enterprise?
First, active matter is condensed matter in a fundamentally new
nonequilibrium regime: (i) The energy input takes place directly at the scale
of each active particle, and is thus homogeneously distributed through the bulk
of the system, unlike sheared fluids or three-dimensional \textit{bulk}
granular matter, where the forcing is applied at the boundaries. (ii)
Self-propelled motion, unlike sedimentation, is force-free: the forces that
particle and fluid exert on each other cancel. (iii) The direction of
self-propelled motion is set by the orientation of the particle itself, not
fixed by an external field.  Indeed, these can be taken as the practical
defining properties of active matter.  A comprehensive theory of this
ubiquitous type of condensed matter is a natural imperative for the physicist,
and should yield a catalogue of the generic behaviours, such as nonequilibrium
phases and phase transitions, the nature of correlations and response, and
characteristic instabilities. Second, therefore, the generic
tendencies emerging from the theory of active matter, unless suppressed by
specific mechanisms, must arise in vivo, which is why biologists should care
about it.  Last, if we can understand active matter, perhaps we can
manufacture faithful imitations of it in chemomechanical systems without 
components of biological origin \cite{activecolloid1,sanorod}

The reader should keep in mind that theories of active matter were formulated
not in response to a specific puzzle posed by experiments but rather to
incorporate living, metabolizing, spontaneously moving matter into the
condensed-matter fold. This was done through minimal models whose consequences
are relatively easy for theoreticians to work out. Natural realizations of
living matter are far from minimal, so comparisons of active-matter theory with
experiment are likely to be qualitative until well-controlled model systems are
devised.  

\subsection{Prehistory}
\label{prehistory}
Although this is not a historical review, it must be stated here 
that the active-matter idea is not new.
In a prescient article, Finlayson and Scriven \cite{finscriv} argued that biological matter could display hydrodynamic instabilities driven by stresses arising from metabolic activity in the bulk of a fluid. 
Unlike in this review, they build active stress-tensor contributions from gradients of existing scalars such as concentration and temperature. They point out that the field poses ``challenges to theoretical rheology'' and is ``virgin territory for experimental research in the physics of complex fluids''. They conclude that ``continuum.. 
analysis... of instability by active stress is probably .... destined to play a large part in 
understanding some of the engines operating at the cellular level in living systems''. The developments surveyed in this article prove them right.   

\subsection{Structure of this article} 
\label{structuresub}
The body of this review consists of several sections each
treating a specific example of an active-matter system. 
In Section \ref{drypolar} the simplest models of moving flocks, without
solvent flow, are introduced and some key properties discussed. Flocks without macroscopic polarity and hence no migration velocity are surveyed in section \ref{dryapolar}. The theoretical framework for collective self-propulsion in a fluid medium is found in Section \ref{activesusp}, with a discussion of general symmetry-based approaches, and the active mechanics of the cytoskeleton in Section 
\ref{cytohydro}. Non-biological self-driven particles
are the subject of section \ref{artificial}. 
The article closes with Section \ref{outlook} which summarizes the achievements and limitations of the approach adopted here to the modelling of active matter and attempts to set a course for future experiments and theory in this vital and rapidly evolving field.      

\section{POLAR FLOCKS ON A SUBSTRATE}
\label{drypolar}
Arrows are \textit{polar} objects, uniform cylinders are \textit{apolar}. A flock moving in one direction has polar order, whereas equal numbers of ants moving from south to north and from north to south are macrscopically apolar even though the individual ants are polar. We begin by discussing polar ordered states. 
\subsection{The Vicsek model}
\label{vicsekmodel} 
In the simplest models \cite{reynolds,vicsek1} of ordering
transitions of active particles, the medium through which the
particles move is treated as an inert substrate. Each particle carries a
velocity vector of fixed magnitude, which it rotates to point parallel to the
mean of its neighbours' velocities, with an angular tolerance $\eta$. It
then takes a small step in the direction of the updated vector. Like the
continuous-spin magnets that they resemble, the Vicsek family of models display
a well-defined phase transition from a disordered phase to a coherent flock as
$\eta $ is decreased or the number density is increased
\cite{vicsek1,gregoirechate1,gregoirechate2}. \textit{Un}like planar spins
\cite{mw,hohen}, flocking models show true long-range order even in two
dimensions. How they do so was shown by Toner and Tu \cite{tonertusr,tonertu,tonertuprl95},
whose continuum field-theoretic approach, which amounts to a coarse-graining of
the Vicsek rules, we sketch below.

\subsection{Toner-Tu field theory} 
\label{tonertutheory} 
A coarse-grained dynamical description works with ``slow variables'' \cite{mpp}, whose relaxation times increase unboundedly with increasing wavelength. For the Vicsek model these are 
\cite{tonertu,tonertuprl95} the number density field $c$, since total number is conserved, and the velocity field ${\bf p}$, whose fluctuations transverse to the mean ordering direction are the spin-waves or Nambu-Goldstone \cite{nambu,goldstone} modes of the flock \cite{ampflucfoot}. With the slow variables in hand, Toner and Tu \cite{tonertu,tonertuprl95} write down the equations of motion, incorporating the physics of the Vicsek model, on general grounds of symmetry. 

First, the dynamics of ${\bf p}$ is described in the equation below: 
\beq 
\label{vecop}
\partial_t {\bf p} + \lambda {\bf p} \cdot \nabla {\bf p} +... = 
(\alpha - \beta {\bf p} \cdot {\bf p}) {\bf p} + \Gamma \nabla \nabla {\bf p}     - \nabla P(c) + {\bf f} 
\eeq 
Ignoring all gradients in Equation \ref{vecop}, 
a phase transition from the isotropic
state ${\bf p} = 0$ to an ordered flock with $|\langle {\bf p}\rangle| \simeq
\sqrt{\alpha / \beta}$, spontaneously breaking rotation invariance, should occur as the control parameter $\alpha$ is tuned from negative to sufficiently large positive
values; $\Gamma$, in general a tensor, controls the elastic \cite{elastviscfoot} restoring forces to distortions in the ordered phase. The random error $\eta$ of the Vicsek update rule is summarized in a nonconserving, Gaussian noise $\mathbf{f}$, uncorrelated in space and time. The crucial difference from traditional models of dynamic critical phenomena \cite{hh} lies in the Navier-Stokes-like $\lambda$ term \cite{galinvfoot}, which says that distortions in ${\bf p}$ are advected by ${\bf p}$, because ${\bf p}$ is not only an order parameter but a velocity. $P(c)$, a general increasing function of concentration, then embodies the equation of state for the pressure. 
Second, c evolves through the continuity equation
\beq
\label{conteqgen}
\partial_t c + \nabla \cdot c {\bf p} = 0, 
\eeq
because $\mathbf{p}$ is the velocity field of the particles. 
Equations \ref{vecop} and \ref{conteqgen} can be viewed as the dynamics of a fluid with a preferred speed relative to a background medium, or of a magnet whose spin is a velocity. In the second interpretation, the nonequilibrium character of the Toner-Tu model enters only through the advective $\lambda$ term in Equation \ref{vecop} and the current in Equation \ref{conteqgen}. The right-hand side of Equation \ref{vecop} can be written as $-\delta F / \delta {\bf p}$ plus noise, with a free-energy functional 
\beq 
\label{polarfree}
F[{\bf p}] = \int d^dr [-{\alpha \over 2}{\bf p} \cdot {\bf p} + {\beta \over 4} ({\bf p} \cdot {\bf p})^2 + {{\Gamma} \over 2} \nabla {\bf p} \nabla {\bf p} + {\bf p} \cdot \nabla P(c)].  
\eeq
of Ginzburg-Landau form, in which the ``pressure gradient'' $\nabla P(c)$ appears as an orienting field for ${\bf p}$. In the absence of $\lambda$, the average dynamics is downhill in $F$. A flock governed by Equations \ref{vecop}-\ref{polarfree} is a strange blend of magnet and fluid. Let us rediscover some of its unique properties \cite{tonertu,tonertuprl95}, all consequences of the fact that the order parameter is itself a velocity. 

\subsection{Sound modes, giant fluctuations, and $2d$ long-range order}
\label{polarflockprops}
\textit{Sound modes}: Let us linearize Equations \ref{vecop}, \ref{conteqgen} about a uniform ordered state with $|\langle {\bf p}\rangle| = p_0$, $c = c_0$, denote directions along and transverse to $\langle{\bf p}\rangle$ by $||$ and $\perp$, and see how small disturbances $\delta c$, $\delta {\bf p} \equiv  (\delta {\bf p}_{\perp}, \delta p_{||})$ travel. As in Heisenberg or XY magnets \cite{chailub}, we can see from Equation \ref{vecop} that $\delta p_{||}$ relaxes rapidly, so that on long timescales $\delta {\bf p} \simeq \delta {\bf p}_{\perp}$. If we take $\lambda =1$ in Equation \ref{vecop}, a shift to a frame moving with speed $p_0$ eliminates advective terms from Equations \ref{vecop} and \ref{conteqgen} in this linearized analysis without losing the essential physics. To leading order in gradients, and ignoring the noise, Equations \ref{vecop} and \ref{conteqgen} then become 
\beq
\label{conceomflock}
\partial_t \delta c = - c_0 \nabla_{\perp} \cdot \delta {\bf p}_{\perp}, 
\eeq
\beq
\label{poleomflock}
\partial_t \delta {\bf p}_{\perp} = -P'(c_0) \nabla_{\perp} \delta c
\eeq
leading to propagating modes with frequency 
\beq
\label{flockdisp}
\omega = \pm\sqrt{c_0P'(c_0)}q_{\perp}
\eeq
at wavevector ${\bf q} = ({\bf q}_{\perp}, q_{||})$. No, we haven't rediscovered normal 
sound waves. Long-wavelength sound in a fluid propagates because of momentum conservation, which a flock lacks. The ``sound modes'' in a flock are a consequence of spontaneously broken rotation invariance: they propagate at long wavelength because $\delta {\bf p}_{\perp}$ is not damped at zeroth order in wavenumber. 

\textit{Giant number fluctuations}: $\delta {\bf p}_{\perp}$ is the Nambu-Goldstone mode \cite{nambu,goldstone} of broken rotation invariance, easily excited and slow to decay at small wavenumber $q$. Hence when noise is included the steady-state variance $\langle |\delta p_{\perp {\bf q}}|^2 \rangle$ of the $q$th fourier component should diverge at small wavenumber $q$, like spin waves in an XY magnet \cite{chailub} or director fluctuations in a nematic \cite{degp}. But Equations \ref{conceomflock}-\ref{poleomflock} imply that, for frequencies and wavenumbers related by Equation \ref{flockdisp}, $\delta c \sim \sqrt{c_0/P'(c_0)}\delta p_{\perp}$. Therefore, the variance of $\delta c$ should also diverge at small $q$. In real space this means that for regions with $N$ particles on average, the variance in the number grows faster than $N$: a flock should display giant number fluctuations. 

\textit{Long-range order in $2d$}: Equations \ref{vecop} and \ref{conteqgen} when linearized yield 
$\langle |\delta p_{\perp {\bf q}}|^2 \rangle \sim 1/q^2$. Taken literally this would mean only quasi-long-range order in dimension $d=2$, by analogy with the XY model. However, nonlinearities in Equations \ref{vecop} and \ref{conceomflock} are stongly relevant in $d=2$.  
and lead to $\langle |\delta p_{\perp {\bf q}}|^2 \rangle$ diverging more slowly than $ \sim 1/q^2$ for most directions of ${\bf q}$, thus preserving long-range order \cite{tonertu,tonertuprl95}. Here is a qualitative explanation of how this happens. Consider a flock with $\langle {\bf p} \rangle = p_0 \hat{\bf z}$ at a given instant. Because ${\bf p}$ is a velocity, a long wavelength fluctuation $\delta {\bf p}_{\perp}$ will allow creatures in the flock to exert their orienting influence on regions out to a distance $\delta p_{\perp} t$ in a time $t$. This effective enhancement of the range of influence ultimately leads \cite{tonertu,tonertuprl95,tutonerulm} to a suppression of orientational fluctuations on large scales, and thus to long-range order in $d=2$. The effect is absent in $d=1$, where rotations are a discrete symmetry and there \textit{is} no $\delta {\bf p}_{\perp}$. We will return to this point in section \ref{1dflocks}. 

I have not discussed microscopic derivations of the Toner-Tu equations \cite{bertinetal,shradhathesis}, dynamics near the onset of polar ordering  \cite{aparnarods,smabmcm}, and the application of flocking models \cite{leekardar,gautam} to motor-microtubule extracts \cite{nedelec}. 

\subsection{Tests of vector flocking models} 
\label{drypolartests} 
Two types of tests of the physicist's approach to flocking are required. Experiments on real flocks are of course the real test. However, it is equally important to check the predictions of coarse-grained theories against computer experiments on microscopic models. We discuss both below, although measurements of the phase-transition variety on real flocks are scarce. 

\subsubsection{Scaling in the ordered phase -- simulations}
\label{orderedscaling}
Quantitative agreement has been found between numerical ``experiments'' on microscopic models and the predictions of the coarse-grained theory, including long-range order in $d=2$, the form of the propagating modes, anomalous density fluctuations and superdiffusion of tagged particles \cite{tonertu,tonertusr,gregoirechate1}. I am not aware of any laboratory or field-based attempts to measure ordered-phase correlation functions for flocks on a substrate. 

\subsubsection{Nature of the flocking transition -- simulations} 
\label{orderoftrans} 
Although some studies \cite{vicsek1,gonci,aldana} claim to see a continuous onset of the ordered phase, as a mean-field treatment of Equation \ref{vecop} or analogy to continuous-spin magnets would suggest, very large-scale numerical studies of the Vicsek model and its relatives show that the transition is in general a discontinuous one, characterized by a complicated coexistence 
\cite{gregoirechate1,gregoirechate2,chateetal2007}. See also Ref. \cite{kulinskii}.  

\subsubsection{On one-dimensional flocks} 
\label{1dflocks} 
In numerical studies of the Vicsek model in one space dimension, long domains of coherently moving particles do appear at low noise and high density. It is occasionally claimed \cite{vicsek1,vicsekcsirokbarabasi} that this is a true phase transition in the limit of infinite system size, as occurs in Vicsek models in dimension $d \geq 2$. 
Recall that the $1d$ Ising model fails to order because of the proliferation of kinks \cite{chailub}. Does some magic suppress kinks, i.e., velocity flips, in $1d$ flocks? Raymond and Evans \cite{raymond_evans} estimate the lifetime $\tau(L)$ of a $1d$ flock of length $L$ as the mean time for the appearance of a kink. They show that if $\tau(L)$ is to grow at least as a power of $L$, the number of particles per site must grow, artificially, at least as $\log L$. Accepting the constraint of a finite number density yields a finite lifetime and correlation length for the would-be flock. Ultimately, nothing saves $1d$ flocks from the fate of the $1d$ Ising model because flocking in $1d$ breaks a \textit{discrete} symmetry and is thus not accompanied by Nambu-Goldstone modes whose long range is responsible for the rescue act in $d=2$. 

Nevertheless, lowering the noise strength or raising the density at fixed $L$ will eventually yield a system with correlation length larger than $L$, and a bistable time-series of the flocking order parameter, i.e., a finite transient flock with a large lifetime. Precisely this phenomenon has been seen in experiments, to which we turn next. 

\subsubsection{Real experiments on flocks on a substrate}
\label{drypolarexpts} 

\begin{enumerate}
\item Couzin \cite{couzin1d} studies the one-dimensional flocking of desert locusts on an annular track. The onset of coherent motion seen in the experiments as the number of locusts is increased, and the bistable time-series switching between clockwise and counterclockwise circuits of the track are well described by the $1d$ Vicsek model \cite{vicsek1}, to which the paper makes a comparison. One must not of course conclude -- see Reference \cite{raymond_evans} and section \ref{1dflocks} -- that there is a true phase transition here. All the same, the experiment demonstrates convincingly the relevance of simple flocking models to an understanding the behaviour of real organisms. 

\item 
Szab\'{o} {\etal} \cite{vicsekkerato} created a polar flock in a petri dish by using motile keratocytes extracted from fish scales. As the cells divide and the areal density of cells crosses a threshold,  there is a well-defined onset of a state of macroscopic order in the orientations and velocities of the keratocytes, that can plausibly be interpreted as a flocking phase transition, which should be governed by the local physics of the models we have been discussing, as the cells interact only by contact. The data quality makes it impossible to draw conclusions regarding the order of the transition. The authors make further inferences on the transition from a computational model, not from data. Despite limitations the experiment is valuable as a realization of the flocking transition, with potential for higher-precision studies including finite-size scaling. 

\item There is a surprising link between vibrated granular monolayers and self-propulsion \cite{sanorod,sradititoner,kudpre2003,tonertusr,aransonrmp}.  Polar rods \cite{sanorod}, placed on a vertically vibrated horizontal linear track, move on average with small end forward. Systematic unidirectional motion of apolar rods \cite{kudpre2003} was effected by maintaining them permanently tilted, by overfilling an annular channel with rods. The propulsion mechanism \cite{volfson,dorbolo,aransonrmp} is illustrated in the cartoon Fig. \ref{shakenrod}.
\begin{figure}
\centerline{\psfig{figure=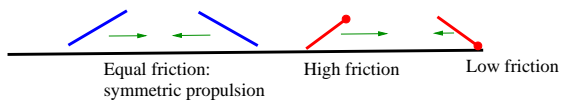,height=3pc}}
\caption{A rod landing after being tossed up will in general be impelled away from the end that makes first contact with the surface. If the two ends differ in weight, geometry or friction, the rod will be propelled towards one end.}
\label{shakenrod}
\end{figure}
The reader is encouraged to check that this system satisfies all three defining properties of active matter in section \ref{activematerialsub}. Published studies so far \cite{kudrolliprl100} of polar rods at high coverage on a vibrated surface report order and anomalous density fluctuations on a local scale, and an abrupt flight to the boundary when the vibration amplitude crosses a threshold, but not a macroscopically ordered flock. The case of apolar rods \cite{vjnmsrscience,vjnmsrjstat} is discussed in section \ref{subsub:dryapolarexps}.
\end{enumerate}

\section{FLOCKS THAT GO NOWHERE ON A SUBSTRATE} 
\label{dryapolar} 
\subsection{The origin of giant number fluctuations in an active nematic}
\label{activecurr}
At least one example is known \cite{gruler} of living, pulsating elongated cells forming a state with spontaneous, uniaxial orientational order with a macroscopic axis $\hat{\bf n}$ with $\hat{\bf n} \to - \hat{\bf n}$ symmetry. This is the spatial symmetry of nematic liquid crystals, and it guarantees that the mean macroscopic velocity of such a ``flock'' is zero. And yet active nematics are profoundly different from dead ones. In particular, here as in polar flocks, giant number fluctuations are predicted \cite{sradititoner} -- even though the order parameter for the nematic is \textit{not} a velocity. Below is an explanation of how it works. 

\begin{figure}
\centerline{\psfig{figure=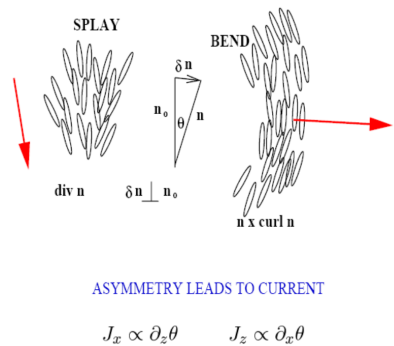,height=12pc}}
\caption{In an apolar nematic, curvature confers local polarity and hence local motion.}
\label{curvcurrfig}
\end{figure}

In the presence of noise, long wavelength fluctuations of the axis of orientation -- the broken-symmetry mode -- are abundant. 
Working in two dimensions for simplicity, where distortions are described by a single angle $\theta$, we see (Fig. \ref{curvcurrfig}) that bend or splay produces a \textit{polar} configuration. 
The absence of time-reversal invariance in a driven state, together with the polarity, means \cite{curie} 
that particles in the curved region must drift in a direction defined by the curvature as pictured in 
Fig. \ref{curvcurrfig}, leading to a current of particles ${\bf J}$ with 
$ J_x \propto  \partial_z \theta, \, J_z \propto \partial_x \theta$ in two dimensions. 
In steady state, this active flux must be balanced on average by restoring diffusive currents $\sim \nabla c$. Thus fluctuations in $c$ and $\theta$ must be of the same order. Orientational fluctuations, being a broken-symmetry mode, should have a variance of order $1/q^2$ at wavenumber $q$. Therefore, number density fluctuations $\langle |c_{\bf q}|^2 \rangle \sim 1/q^2$ as well which means the number fluctuations in regions containing $N$ particles on average have a standard deviation 
\beq
\label{GNF}
\Delta N \propto N^{1/2 + 1/d} \sim N \, \mbox{for} \, d = 2
\eeq 
for large $N$. Unlike for polar flocks \cite{tonertu}, these results of the linearized theory for apolar systems survive \textit{quantitatively} when nonlinearities are included \cite{shradhaRG,shradhathesis}.   

\subsection{Tests of apolar flock theories: experiments and simulations}
\label{apolarsim}
\subsubsection{Particle models for active apolar order} 
\label{chatemodel}
The first test of the existence of giant number fluctuations (GNFs) in active nematics was a computer experiment on a clever apolar generalization \cite{chateginellimontagne} 
of the two-dimensional Vicsek model described in section \ref{vicsekmodel}. Each particle is assigned an \textit{axis}, rather than a unit vector, which it aligns parallel to the mean of its neighbours' axes, subject to a small angular noise. The particle then takes a small step preferentially along the axis, forward or back. Since the orientation variable and the moves are all defined modulo $\pi$, the model has no polarity. As noise or mean interparticle separation is decreased, a continuous transition is seen from an isotropic state to a phase with quasilong-range nematic order of the particle axes as in equilibrium \cite{degp,veerman}. However, the density fluctuations are giant: $\Delta N$ grew far more rapidly than $N$ and could be fitted to a form $a\sqrt{N} + b N$ in precise agreement with the predictions \cite{sradititoner} whose derivation was sketched above. 
Real-space snapshots showed strong transient inhomogeneity, with a banded structure. If the anisotropic, detailed-balance-violating move is replaced by an isotropically distributed random step, the nematic phase survives, but with normal number fluctuations, because detailed balance now holds. S. Mishra \cite{shradhathesis} has shown that coarse-graining the model of Chat\'e {\etal} \cite{chateginellimontagne} yields the equations of motion of Ref. \cite{sradititoner}, with the active current as in section \ref{activecurr}. Further numerical studies on a particle model \cite{shradhasr} show a close connection between the giant number fluctuations of an active nematic and the phenomenon of fluctuation-dominated phase ordering \cite{dasbarma}. 

\subsubsection{Experiments on apolar flocks on a substrate} 
\label{subsub:dryapolarexps} 
\begin{enumerate}
\item \textit{Melanocyte nematics:} Living melanocytes, the cells that spread pigment in skin, have been shown \cite{gruler} to form apolar, nematic order in vitro as concentration is increased. The resulting $2d$ nematic phase has large regions of oriented cells with occasional point topological defects of strength $-1/2$ which reinforce the nematic interpretation \cite{degp,gruler}. Although the authors of \cite{gruler} emphasize that their systems are very far from thermal equilibrium the analysis they bring to bear on the problem, especially in the apolar case, is largely inherited from the equilibrium theory of liquid crystals.  
The focus in \cite{gruler} is on modelling cell orientations, in a population-averaged description which cannot resolve spatial variations in the density. Consistent with this limitation, their experiments do not look at the behaviour of density fluctuations. The purpose of this discussion is to urge experimenters to take another look at the melanocyte system, especially as the strange properties of active nematics \cite{sradititoner,chateginellimontagne,shradhasr} remain to be checked in a biological experiment. 

\item \textit{Active apolar states in granular matter:}
\begin{figure}
\centerline{\psfig{figure=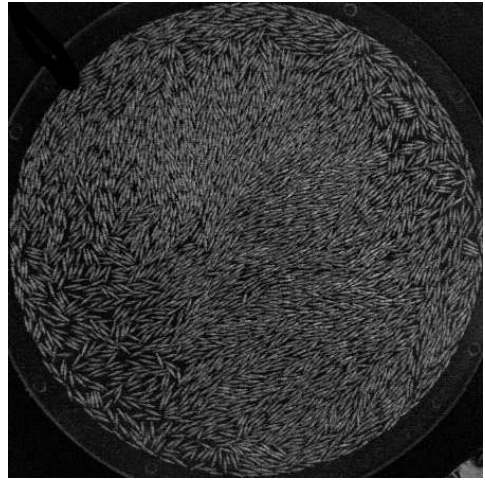,height=15pc}}
\caption{Snapshot from the active nematic phase of vibrated copper-wire segments. Figure credit: Vijay Narayan}
\label{nemimagelow}
\end{figure}

Narayan {\etal} \cite{vjnmsrscience,vjnmsrjstat} carried out a systematic study of liquid-crystalline order in a vertically agitated monolayer of copper-wire segments. In particular, they measured number density fluctuations in the nematic and the isotropic phases. Their findings \cite{vjnmsrscience} confirmed the predictions \cite{sradititoner} of giant number fluctuations in the nematic phase. A typical nematic configuration and a plot of standard deviation \textit{vs} mean number are shown in Figs. \ref{nemimagelow}
and \ref{GNFlow}. 
\begin{figure}
\centerline{\psfig{figure=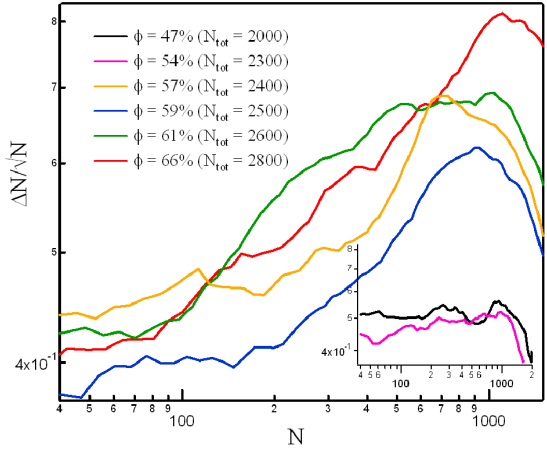,height=15pc}}
\caption{Giant number fluctuations in experiments on the copper-wire active nematic: standard deviation $\Delta N$ scaled by $\sqrt{N}$ grows with mean number $N$. Inset: no such growth in the isotropic phase. Figure credit: Vijay Narayan}
\label{GNFlow}
\end{figure}
 An additional consequence of the theory that the autocorrelation of the local density should decay as $-\log t$ over a large time range \cite{sradititoner} was also confirmed \cite{vjnmsrscience}. Lastly, single-particle tracking showed that the rods moved preferentially along their length just like the apolar flockers in the model of Chat\'{e} {\etal} 
\cite{chateginellimontagne}. 

Other experiments of interest, on chiral structures in active granular matter, include the work of 
Tsai {\etal} \cite{fangfu} and of Blair {\etal} \cite{kudpre2003}. 
The measurements on granular matter reported above are among the few quantitative tests of active-matter theories. That they are not carried out on living systems only serves to reinforce the universality of the approach reviewed in these pages. 

\end{enumerate}
 
\section{SWIMMING ORGANISMS AND THE CYTOSKELETON}
\label{activesusp} 
\subsection{The equations of active hydrodynamics}
\label{activehydrosection}

\begin{figure}
\centerline{\psfig{figure=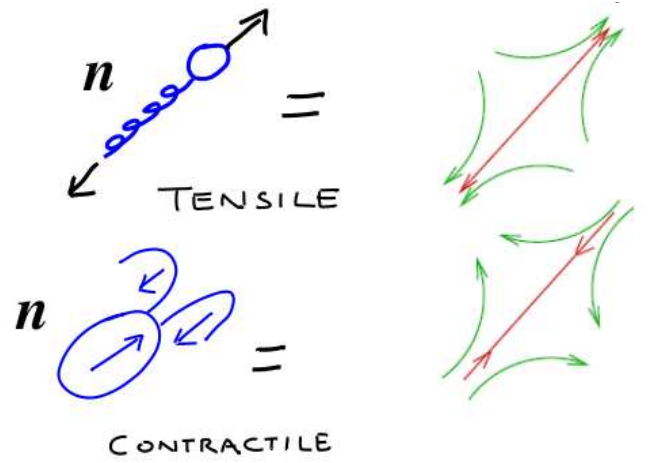,height=15pc}}
\caption{The flows around a swimming bacterium (top) and an algal cell (bottom): bacteria are extensile, and algae contractile, force dipoles.}
\label{bactalg}
\end{figure}

The collective motion of swimming organisms can be understood by constructing the coupled dynamical equations for the swimmer concentrations and orientations, and a generalized Navier-Stokes equation for the velocity field ${\bf u}$ of the suspension. Self-propelling activity enters via force densities in the Navier-Stokes equation. These have no monopole moment, since the mutual forces of swimmer and fluid cancel by Newton's 3rd Law \cite{pedleykessler}. The minimal model of an active particle in a fluid is therefore a permanent force dipole (see Fig. \ref{bactalg}), whose strength $W$ is positive for extensile and negative for contractile swimmers [See also the modelling of ion pumps in a model membrane by Manneville {\etal} \cite{pumplong}.] 
Associating a local orientational order parameter field ${\bf p}({\bf r},t)$ with the vectorial orientation of the swimmers in a coarse-graining cell around ${\bf r}$ at time $t$, a straightforward calculation \cite{aditisr} shows, to leading order in a gradient expasion, that a collection of active particles at concentration $c({\bf r},t)$ has an active force density ${\bf F}_a({\bf r}, t) = -\nabla \cdot \sigma^a$ with an active stress 
\bea
\label{sigact}
\sigma^a &=& W c({\bf r},t) {\bf p} {\bf p} \nonumber \\ 
&=& W c({\bf r},t) \bsf{Q}({\bf r},t) + {W \over 3} c p^2 \bsf{I},  
\eea
where $\bsf{I}$ is the unit tensor, and $\bsf{Q}$, the trace-free part of ${\bf p} {\bf p}$, determines the deviatoric stress which is all that concerns us in the incompressible limit $\nabla \cdot {\bf u} = 0$. 
Activity thus produces flow, and flow in turn reorients the principal axes of rodlike particles preferentially along the extensional axis (and platelike along the compressional axis), as it would in non-living liquid crystals as well. The resulting equations of motion for ${\bf p}$ are obtained by adding flow-orientation couplings to Equation \ref{vecop}:
\beq 
\label{vecopwet}
D_t {\bf p} + \lambda {\bf p} \cdot \nabla {\bf p} + .... = \gamma \bsf{A}\cdot {\bf p} 
- {\delta F \over \delta {\bf p}} + {\bf f},  
\eeq 
where $D_t$ is the time-derivative in a frame locally \textit{comoving} and \textit{corotating} with the fluid, the coefficient $\gamma$ governs the orienting effect of the extensional part of the flow through the term in $\bsf{A} \equiv [\nabla {\bf u} + (\nabla {\bf u})^T]/2$, and the remaining terms are as defined for Equation \ref{vecop}, $F$ being the free-energy of Equation \ref{polarfree}. Inertia, viscosity $\eta$, elastic forces from Equation \ref{polarfree}, and activity compete to determine the dynamics of the suspension velocity field: 
\beq
\label{NSactive}
\rho (\partial_t + {\bf u} \cdot \nabla){\bf u} = -\nabla \cdot (\sigma^a + \sigma^h) -\eta \nabla^2 {\bf u} -\nabla \Pi; \,  \nabla \cdot {\bf u} = 0, 
\eeq
with incompressibility $\nabla \cdot {\bf u} = 0$ enforced by the pressure $\Pi$. 
In Equation \ref{NSactive}, $\sigma^h$ is the contribution to the stress from Equation \ref{polarfree} for the free-energy functional \cite{degp}. Lastly, the
current in the continuity equation for the concentration, in the lab frame, is
$({\bf p} + {\bf u})c$, because we can simply identify the orientation field ${\bf p}$ with the active-particle velocity with respect to the fluid. 

We turn next to the dramatic consequences of this dynamics. 

\subsection{The instability of orientationally ordered active suspensions} 
\label{instabsec} 
\begin{figure}
\centerline{\psfig{figure=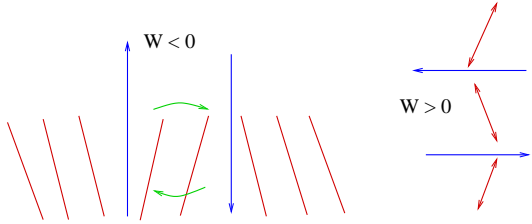,height=7pc}}
\caption{The heart of the generic instability of ordered active filaments: long-wavelength splay produces shear flows that further distort a row of parallel contractile force dipoles (left); similarly bend disrupts extensile filaments (right).}
\label{splaybendinstab}
\end{figure}
Active suspensions with uniform orientational order are in a state of permanent uniaxial tension or compression and are therefore intrinsically unstable \cite{aditisr,voit,srmrnjp}: An individual contractile particle (Fig. \ref{bactalg}) pulls fluid in from both ends along its main axis. In a perfectly ordered, unbounded, parallel collection of contractile particles, these self-generated flows cancel. A long-wavelength splay fluctuation disrupts this delicate balance, resulting in flows upward on one side and downward on the other, as shown in Fig. \ref{splaybendinstab}. The ensuing shear will amplify the rightward tilt of the middle portion, hence the instability. The same argument, \textit{mutatis mutandis}, implies a \textit{bending} instability of extensile filaments \cite{llelast}. A detailed solution \cite{aditisr,voit,srmrnjp} of Equations \ref{vecopwet}, \ref{NSactive} 
bears out this pictorial argument.  

Note that we've assumed instant, global response of the fluid flow, which amounts to the steady Stokesian approximation in which inertia and acceleration are ignored in Equation \ref{NSactive}. In that limit, in an unbounded system with mean concentration $c_0$, the growth rate of the instability must be proportional to the only available inverse timescale, namely, the ratio $W c_0 / \eta$ of active stress to viscosity. In a container with smallest dimension $L$, active stresses compete \cite{srmrnjp}with orientational elasticity from Equation \ref{polarfree}, resulting in a threshold $W c_0 \sim K/L^2$ for the instability. Far beyond threshold, scaling at wavenumber $q$ yields a growth rate 
$(qL)^2 W c_0/ \eta$ for $qL \ll 1$, crossing over to $W c_0 / \eta$ for $qL \gg 1$. The detailed form of the instability depends on the experimental geometry. In Ref. \cite{voit}, for example, it appears as an active version of the Freedericksz transition of nematic liquid crystals \cite{degp}. Novel instability mechanisms recently noted in polar systems include oscillations and banding if a concentration field is included \cite{giomi}, and travelling undulations if a deformable 
free surface is present \cite{sumithra}. Aranson {\etal} \cite{aransonthin} study free-standing active thin films and introduce \textit{ad hoc} a propulsive force simply proportional to the thickness-averaged ${\bf p}$. Such a term can be obtained from the treatment of this section and Ref. \cite{aditisr} if ${\bf p}$ is neither parallel nor perpendicular to the free surface. 
The stabilizing effect of shear has been discussed by Muhuri {\etal} \cite{muhuri}. 

There is detailed evidence for the basic instability in numerical studies of particles self-propelled through a fluid  \cite{shelley1,shelley2,ishikawapedley,ishikawareview,locsei,mehandianott,graham2005prl,ganesh}, who also point out a variety of instabilities in the isotropic phase.  
Studies of the PDEs of active hydrodynamics include Refs. \cite{marendu,marenduetalpre2007,marenduetaljnnfm,llopis,sanoop,wolg} who show that the linear instability leads to complex flow patterns and turbulence driven not by Reynolds number but an active Ericksen number $(W c_0 / \eta) \tau$ where $\tau$ is the collective relaxation time of the $\bsf{Q}$ tensor. 

For small but nonzero Reynolds number Re$_a$ on the length scale $a$ of a particle, the instability arises for wavenumber $q$ with 
\beq
\label{Rerange}
\mbox{Re}_a \ll qa \ll 1,  
\eeq
a range that clearly does not exist for swimmers of macroscopic size and speed.  

\subsection{The effect of activity on viscosity} 
\label{activerheosec}
The enhancement or reduction of viscosity by activity  
\cite{activerheo} in a suspension in the isotropic phase is one of the most robust predictions of active hydrodynamics. 
\begin{figure}
\centerline{\psfig{figure=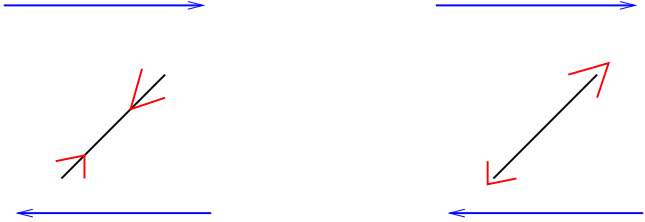,height=7pc}}
\caption{How activity modifies viscosity: Shear orients filaments; the permanent force dipoles pull back, if contractile, and push out, if extensile, on the flow.}
\label{activevisc}
\end{figure}
The mechanism is readily understood, pictorially (see Fig. \ref{activevisc}) or verbally: consider a collection of filaments endowed with contractile or extensile force dipoles, in the isotropic phase but at high enough concentration that collective orientational relaxation is slow. A modest imposed shear flow will then produce appreciable alignment. Contractile filaments will then pull back on, and extensile filaments push out on, the flow that oriented them. The result is a higher stress/rate ratio, that is, a viscosity, for contractile filaments, and lower for extensile filaments, than would be found for dead particles of the same shape and at the same concentration. Upon tuning a parameter such as concentration to increase the collective orientational relaxation time $\tau$, a suspension of contractile filaments should show strong viscoelasticity, like an equilibrium system approaching a glass transition \cite{activerheo,tanniecrisactiverheo} and extensile filaments should show enhanced unstable shear-thinning. Numerical studies \cite{catesfielding} find a complex active rheology including the predicted \cite{activerheo,tanniecrisactiverheo} glassy enhancement of viscosities for contractile systems, and shed valuable light on the role of boundary conditions. Microscopic models for filaments and motors \cite{tanniecrisactiverheo} and swimmers \cite{saintillanexpmech} recover the predictions of Ref. \cite{activerheo}. 

\subsection{Fluctuating active hydrodynamics} 
\label{flucactivehydro}
The enhancement of biological noise by active processes was studied some years ago in the context of membranes with pumps \cite{prostbruinsma1,pumppramana,pumpprl,pumplong,cras,gautammembrane}.  
A natural explanation for the enhanced noise temperature of swimming bacteria \cite{xlwu} was found in the active hydrodynamic framework by including random forces, torques and currents \cite{aditisr,activerheo} which are consistent with conservation laws but unrelated a priori to transport coefficients, because these are systems far from thermal equilibrium. Lau and colleagues \cite{lau1,lau2} model the microrheology of bacterial suspensions in the active hydrodynamic framework with noise. They relate the collective noise to the tumbling of bacteria, and show that Equation \ref{sigact} implies a $t^{-1/2}$ decay of the stress autocorrelation for times shorter than the collective orientational relaxation time $\tau$. They also show, as do Underhill {\etal} \cite{grahamprl2008}, that equal-time correlations of the suspension velocity field on intermediate length scales $r$ should decay as $1/r$. Related predictions arise in work by Golestanian \cite{raminactive} in the \textit{a priori} different context of catalytically self-propelled colloids \cite{activecolloid1}. Active noise in the ordered phase is predicted \cite{abhiketal} to lead to diffusivities that depend on sample thickness, through a mechanism related to the giant fluctuations \cite{tonertu,tonertuprl95,tutonerulm,aditisr,sradititoner} discussed in sections \ref{drypolar} and \ref{activecurr}. 

\subsection{Fish schools} 
\label{fishschoolsec}
Schools of fish \cite{lighthillfish} appear to be  spontaneously ordered phases at high Reynolds number \cite{weihs} and, from Equation \ref{Rerange}, evade the instability of the oriented state of low-Re self-propelled particles. As a first step towards applying active hydrodynamics to macroscopic swimmers, Ref. \cite{aditisr} studied linear perturbations about an ordered steady state of uniform number density and propulsion velocity relative to a quiescent fluid. Ignoring inertia and retaining active stresses and the Navier-Stokes acceleration term in Equation \ref{NSactive}, they found that all modes were propagative, with highly anisotropic speeds of order $\sqrt{W c_0/\rho}$, and showed that the inclusion of fluid flow did not eliminate the giant number fluctuations in polar flocks predicted by Toner and Tu \cite{tonertu}. It would seem worthwhile to test these predictions on fish schools. See Ref. \cite{makris}, to which we return in section \ref{fishsofq} below.  . 

\subsection{Active hydrodynamics of the cytoskeleton}
\label{cytohydro}

The cytoskeleton with its motors, on timescales long compared to the release time of transient crosslinks, is also an active fluid suspension of polar orientable objects \cite{activegel1,activegel2,kruseetalnjp,physrep,hfsp} with macroscopically contractile stresses along the filament axes \cite{takiguchi}. 

\subsubsection{The active-gel picture}
\label{activegelsection}
Hydrodynamic equations for \textit{active gels} \cite{activegel1,activegel2,physrep,hfsp} were formulated with a view to modelling the cytoskeleton, mainly on long timescales when it is fluid. 
The description turns out to be precisely the same as that of Ref. \cite{aditisr}, but the construction uses linear irreversible thermodynamics \cite{degroot,mpp} for a nonequilibrium steady state \textit{near equilibrium}. A linear relation is assumed \cite{activegel1} between ``fluxes'' (including the stress $\sigma_{ij}$ and the rate $r$ of ATP hydrolysis) and ``forces'' including the velocity gradient $\nabla_i u_j$ and the chemical potential difference $\Delta \mu$ between ATP and its reaction products, allowing for a nondiagonal matrix of kinetic coefficients. 
The existence of the polar order parameter ${\bf p}$ makes it possible to construct a scalar 
$\zeta p_i p_j \nabla_i u_j$ contribution to $r$, linear in $\nabla {\bf u}$, where $\zeta$ is a phenomenological parameter. By the (anti)symmetry of reversible kinetic coefficients, this implies a piece $ -\zeta p_i p_j \Delta \mu$ in the stress $\sigma_{ij}$, which we recognize as the active stress of Equation \ref{sigact}. 

Below is some recommended reading on recent applications of active hydrodynamics to the cytoskeleton. Active stresses have been shown \cite{salbreux} to be important in the formation of the contractile ring around the equator of an animal cell about to divide. 
Oscillations in experiments on non-adhering fibroblast cells have been understood through the interplay of active stresses in the cortical actin layer with the entry of calcium through ion channels \cite{pramodosc}. Strong departures from the fluctuation-dissipation theorem (FDT) \cite{fdt} have been seen in a permanently crosslinked actin network studded with myosin motors, and analysed using active hydrodynamics extended to the case of a permanently crosslinked network \cite{mizuno,alexfredegennesissue}. Strongly non-FDT behaviour has been predicted \cite{noriopap} in an active hydrodynamic treatment of a stiff filament in an active medium, and connections drawn to the dynamics of auditory hair cells \cite{hudspeth} and axons \cite{pramod,pramodphysrep}. The idea of actin-based active transport parallel to the cell membrane has been proposed in Ref. \cite{jitumadan}. 
 
\subsubsection{From motors and filaments to active hydrodynamics}
\label{marchettiapproach} 
Kruse and J\"ulicher \cite{krusejulicherprl2000}, motivated by actomyosin phenomena, studied active stresses in one-dimensional microscopic models of motors and filament bundles. In an important series of papers, Liverpool, Marchetti and collaborators \cite{cris1,cris2,tanniecrisactiverheo} have carried out a systematic construction of the coarse-grained equations of motion of active filaments, for systems on a substrate as well as in bulk suspension, starting from a microscopic description of motors gripping filaments and moving along them. 
They obtain equations of the form proposed by purely phenomenological theories \cite{tonertu,leekardar,aditisr,sradititoner,activegel1,gautam}, but each parameter in the coarse-grained theories is traced to a microscopic process involving motors and filaments. An important outcome \cite{cris1,cris2,tanniecrisactiverheo} is phase diagrams including the various well-known instabilities of actin solutions such as bundling, with a connection to experimentally accessible control parameters such as filament concentration and strength of motor activity. The initial approach was motivated by cytoskeletal physics, but more recently \cite{aparnacristinaPNAS} the focus has been on swimmers.

\subsection{Experiment on active suspensions}
\label{suspexpts}

\subsubsection{Viscosity measurements on microbial suspensions}
\label{activeviscexps}\begin{figure}
\centerline{\psfig{figure=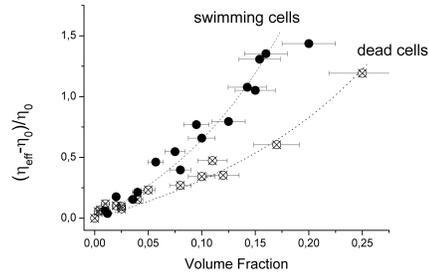,height=10pc}}
\caption{Relative viscosity of suspensions of chlamydomonas (measured at shear rate $5$ 
s$^{-1}$) versus volume fraction $\phi$. Solid symbols: live cell data; crossed symbols: dead cell data. Reprinted figure with permission from S. Rafa\"{\i} {\etal} {\cite{rafai}} (http://prl.aps.org/abstract/PRL/v104/i9/e098102). Copyright 2010 by the American Physical Society.}
\label{rafaifig}
\end{figure}
The prediction \cite{activerheo} that activity alters the viscosity of suspensions, as discussed in section \ref{activesusp}, has recently been tested by {Rafa\"\i} {\etal} \cite{rafai,srcmjc} on the motile alga chlamydomonas, a contractile swimmer, in a conventional rheometer, and by Sokolov and Aranson \cite{sok} on the bacterium Bacillus subtilis, an extensile swimmer, by monitoring the decay of an imposed vortex as well as by measuring the viscous torque on a rotating particle. The idea was to check whether the change in viscosity could be attributed directly to the swimming activity. Fig. \ref{rafaifig} shows the viscosity of the microbe suspension as a function of concentration $\phi$, and compares the values obtained with live and dead cells, for $\phi$ upto 25\%. The dependence of viscosity on $\phi$ was about twice as strong for live as for dead cells. For B. subtilis in \cite{sok}, the viscosity was suppressed by as much as a factor of 7 in some cases. These experiments not only confirm the predictions of Hatwalne {\etal} \cite{activerheo} but also provide strong support for the idea that a medium suffused with active processes such as swimming should be thought of as a distinct type of material. 

\subsubsection{Order and instability of active suspensions}
\begin{figure}
\centerline{\psfig{figure=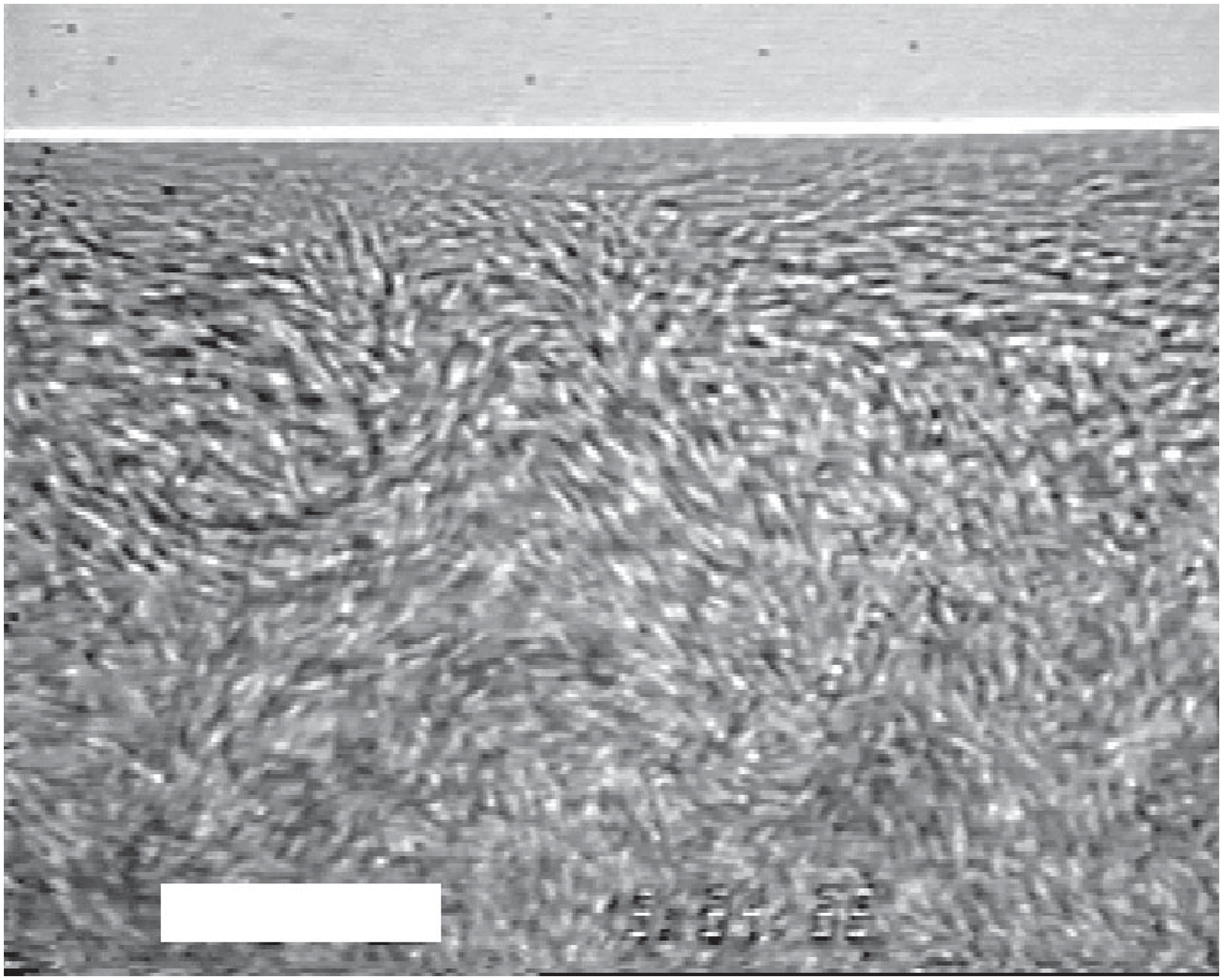,height=10pc}}
\caption{Turbulence at zero Reynolds number in a sitting drop of Bacillus subtilis, viewed from below. Scale bar = 35 $\mu$m. Reprinted figure with permission from C. Dombrowski {\etal}, Phys. Rev. Lett. {\bf 93}, 098103 (2004); \texttt{http://prola.aps.org/abstract/PRL/v93/i9/e098103}. Copyright (2004) by the American Physical Society.}
\label{rayfig}
\end{figure}
There is ample qualitative evidence of the instability of active ordered suspensions predicted in reference \cite{aditisr} and discussed in section \ref{instabsec}. It would appear that collections of bacteria can't swim straight even if they set out to do so \cite{dombrowski,laugareview,sok1}, and that their own active stresses are responsible.  Bacterial turbulence is seen in Fig. \ref{rayfig}. 
The role of instabilities in focusing the bacterial concentration has been stressed by Goldstein and coworkers \cite{dombrowski,cisnerosexpfluids}.


\subsubsection{The structure factor of fish shoals}
\label{fishsofq}
A remarkable underwater acoustic measurement \cite{makris} of the static structure factor of the number density of fish shoals, for wavenumbers from $10^{-4}$ to $10^{-2}$ m$^{-1}$ reveals a power law $\sim k^{-1.5}$, and a rapid wavelike response of fish to perturbations, both of which are consistent with the Toner-Tu \cite{tonertu} work described above. However, to deal with the fact that the fish are in water, a more complete theory of active suspensions at high Reynolds number, or at least in the unsteady-Stokes description, is clearly needed. Power counting on Equation \ref{NSactive} shows that nonlinearities are relevant in dimension $d<4$. An actual renormalization-group calculation for this problem is even more daunting than that for the Toner-Tu model \cite{tonertu} and therefore has not been done. A useful first step would be a comparison of the fish-shoal observations \cite{makris} to the unsteady-Stokes predictions \cite{aditisr} discussed under ``Fish Schools'' in section \ref{activesusp}.  

\section{ARTIFICIAL SELF-DRIVEN PARTICLES}
\label{artificial}
We have already seen one imitation of active matter in section \ref{subsub:dryapolarexps}, namely, a vertically agitated horizontal layer of bits of wire. Several efforts are underway to make colloidal active matter, including beads or rods half covered with platinum and immersed in H$_2$O$_2$ \cite{activecolloid1,activecolloid2}. The platinum catalyzes 2H$_2$O$_2$ $\to$ 2H$_2$O $+$ O$_2$, and the asymmetry leads to directed motion as a result of osmotic gradients. These have been analyzed theoretically in References  \cite{raminetalprl2005,raminactive,ruecknerkapral}. Electro\"osmotic propulsion has been explored by Lammert {\etal} \cite{lammertetal} A magnetically driven artificial flagellum has been demonstrated by Dreyfus {\etal} \cite{dreyfus}. 
The study of self-propelled liquid drops on a surface, a common sight on a hot skillet in the kitchen, is a subfield in itself and cannot be reviewed in any detail here. The driving force is generally evaporation of one or more constituents from the liquid. A couple of recent examples are the work of Thakur {\etal} \cite{snigdha} on nematic drops and Chen {\etal} \cite{yoshi} on oil drops driven by surface tension gradients. It should also be possible to make motors by forcing asymmetric elastic dimers with nonequilibrium noise \cite{vk,adrian}. An intriguing new theoretical development holds out the possibility of using mechanical agitation to make Stokesian swimmers from collections of elastic dimers \cite{bartolo_lauga_shaken}. In all these cases a well-known general principle \cite{curie} seems to work: polarity together with sustained dissipation leads to motility. 

\section{CONCLUSION AND OUTLOOK}
\label{outlook}
What has the active-matter effort achieved? It has opened a new area of fundamental physics -- the hydrodynamics of liquid crystals and suspensions in an entirely novel regime. It is part of a larger process: biology is increasingly concerned with the mechanical forces exerted in living systems, and the active-matter approach provides the framework in which to ask these questions.  
It has made specific predictions, and experimenters are managing to check some of them. 
The impact on biology is already being felt, from the subcellular scale to starling flocks \cite{starflag}. In cell biology, tissue is clearly the next mountain, and some steps on it have already been taken \cite{jfjptissue,boristissue}. The active hydrodynamic framework, suitably adapted, should apply to self-driven particles in more complex media and geometries \cite{durotaxis,safranmechanosens,sandlizard,laugaviscoel,bioconvection,sanoPLoS1}. 

The approach to active matter advocated in this
review shares an important weakness with all coarse-grained theories: an
abundance of phenomenological parameters. If thermal Brownian motion and active
processes are present in comparable magnitude, a clear distinction can probably
be made between parameters determined primarily by equilibrium physics and
arising strictly from activity. However this nice separation is lost if noise
and coefficients themselves are functions of activity \cite{srcmjc}, as they probably are in
bacteria and certainly are in granular matter. This limits somewhat the
predictive power of coarse-grained theories. This is a limitation we're used to
in equilibrium physics: Ginzburg-Landau free-energy functionals predict
temperature dependences best near continuous phase transitions, not over an
entire phase. In addition, living systems are finite in space and time and thus not ideally suited for gradient-expanded approaches. We must admit these limitations and accept humbly the need for microscopic theories.  

There are two imperatives if the field is to make a real contribution to biology. Future theory work must aim to integrate the combined effects of growth, cross-linking, treadmilling, multiple species of filament, and the coupling of the cytoskeleton to the cell membrane \cite{jitumadan}. 
Future experimental work must aim for quantitative tests of existing predictions of active-hydrodynamics, on controlled experimental systems. 

\section{ACKNOWLEDGEMENTS}
\label{ack}
Support from CEFIPRA project 3504-2 and from the DST, India through the CCMT, SR/S4/MS:419/07 and a J C Bose Fellowship are gratefully acknowledged. I thank Salima {Rafa\"{\i}} for a useful discussion and the use of Fig. \ref{rafaifig}, and Ray Goldstein for Fig. \ref{rayfig}.


\begin{thebibliography}{200}
\bibitem{philnelsonbook} Nelson P 2003. Biological Physics: Energy, Information, Life
(Freeman, New York, 2003.

\bibitem{robphillipsbook} Phillips R 2008. Physical Biology of the Cell (Garland Science)

\bibitem{alberts} Alberts B, Johnson A, Lewis J, Raff M, Roberts K, Walter P. 
2002 Molecular Biology of the Cell 4th edn (New York: Garland)

\bibitem{kolomeisky_fisher} Kolomeisky AB, Fisher ME 2007. Annu. Rev. Phys.
Chem. {\bf 58}:675-695

\bibitem{frey08} Mobilia M, Reichenbach T, Hinsch H, Franosch T, Frey E. 2008. 
Banach Center Publications {\bf 80}:101-120; {\tt http://arxiv.org/abs/cond-mat/0612516} 

\bibitem{debchow} Chowdhury D 2008. {\tt http://arxiv.org/abs/0807.2731}

\bibitem{tonertusr} Toner J, Tu Y, Ramaswamy S 2005. Ann. Phys. {\bf 318}:170-244 

\bibitem{physrep} J\"ulicher F, Kruse K, Prost J, Joanny J-F 2007. Phys. Rep. {\bf 449}:3-28    

\bibitem{reynolds} Reynolds C 1987. Computer Graphics {\bf 21}: 25

\bibitem{vicsek1} Vicsek T, Czir\'ok A, Ben-Jacob E, Cohen I, Shochet O. 1995. 
Phys Rev Lett {\bf 75}:1226-29 

\bibitem{tonertu} Toner J, Tu Y. 1998. Phys. Rev. E 58:4828–58
                                                                                           
\bibitem{tonertuprl95} Toner J, Tu Y 1995. Phys. Rev. Lett. 75:4326-29
                                                                                          
\bibitem{tutonerulm} Tu Y, Toner J, Ulm M 1998. Phys. Rev. Lett. 80:4819-22

\bibitem{sradititoner} Ramaswamy S, Simha R A, Toner J. 2003. Europhys, Lett. {\bf 62}:196

\bibitem{vjnmsrscience} Narayan V, Menon N, Ramaswamy S 2007. Science {\bf 317}:105

\bibitem{aditisr} Simha RA and Ramaswamy S 2002. Phys Rev Lett {\bf 89}:058101   

\bibitem{degp} de Gennes PG and Prost J 1993. The Physics of Liquid Crystals, second ed., Clarendon Press, Oxford. 

\bibitem{kaeswebsite} Gerdelmann J, Pawlizak S 2009. Online review of cytoskeletal physics at 
http://www.uni-leipzig.de/~pwm/web/introduction-cytoskeleton.htm

\bibitem{activegel1} Kruse K, Joanny J-F, J\"ulicher F, Prost J, Sekimoto K. 2004 
Phys. Rev. Lett. {\bf 92}:078101

\bibitem{activegel2} Kruse K, Joanny J-F, J\"ulicher F, Prost J, Sekimoto K. 2005. 
Eur. Phys. J. E {\bf 16}:5

\bibitem{cellmaterial} Kasza KE, Rowat AC, Liu J, Angelini TE, Brangwynne CP, et al. 2007. 
Curr Opin Cell Biol {\bf 19}:101-107 

\bibitem{activecolloid1} Paxton WF, Kistler KC, Olmeda CC, Sen A, St. Angelo SK, et al. 2004. 
J. Am. Chem. Soc. {\bf 126}:424

\bibitem{sanorod} Yamada D, Hondou T, Sano M 2003. Phys. Rev. E {\bf 67}:040301.

\bibitem{finscriv} Finlayson BA and Scriven LE 1969. Proc R Soc Lond A {\bf 310}:183 

\bibitem{gregoirechate1} Gr\'egoire G and Chat\'e H 2004. Phys Rev Lett {\bf 92}:025702 

\bibitem{gregoirechate2} Chat\'e H, Ginelli F, Gr\'egoire G and Raynaud F 2008. 
Phys Rev E {\bf 77}:046113 

\bibitem{mw} Mermin ND and Wagner H 1966. Phys Rev Lett {\bf 17}:1133  

\bibitem{hohen} Hohenberg PC 1967. Phys. Rev. {\bf 158}:383 

\bibitem{mpp} Martin PC, Parodi O, Pershan PS 1972. Phys. Rev. A {\bf 6}:2401 

\bibitem{nambu} Nambu Y 1960. Phys Rev {\bf 117}:648

\bibitem{goldstone} Goldstone J 1961. Nuovo Cim. {\bf 19}:154 

\bibitem{ampflucfoot} Fluctuations in the magnitude $|{\bf p}|$ are slow only at a continuous ordering transition.

\bibitem{elastviscfoot} A matter of interpretation: if ${\bf p}$ is regarded as a velocity, $\Gamma$ is like a viscosity.

\bibitem{hh} Hohenberg PC, Halperin BI 1977. Rev. Mod. Phys. {\bf 49}: 435 

\bibitem{galinvfoot} Lacking Galilean invariance, we cannot impose $\lambda = 1$. The ellipsis denotes the two other allowed terms \cite{tonertu,tonertuprl95,cris2} with one $\nabla$ and two factors of ${\bf p}$.

\bibitem{cris2} Ahmadi A, Marchetti MC, Liverpool TB. 2006. Phys Rev E {\bf 74}:061913 

\bibitem{chailub} Chaikin PM and Lubensky TC 1998. 
Principles of Condensed Matter Physics, Cambridge University Press, New Delhi 

\bibitem{bertinetal} Bertin E, Droz M, Gr\'egoire G 2006. Phys. Rev. E {\bf 74}:022101 1-4

\bibitem{shradhathesis} Mishra S 2009. Dynamics, order and fluctuations in active nematics: 
numerical and theoretical studies. Ph D Thesis, Indian Institute of Science


\bibitem{aparnarods} Baskaran A, Marchetti MC 2008. Phys. Rev. Lett. {\bf 101}:268101.

\bibitem{smabmcm} Mishra S, Baskaran A, Marchetti MC. 2010. arXiv:1001.3334


\bibitem{leekardar} Lee HY, Kardar M 2001. Phys. Rev. E {\bf 64}:056113 

\bibitem{gautam} Sankararaman S {\etal} 2004. Phys. Rev. E {\bf 70}:031905

\bibitem{nedelec} Nedelec FJ, Surrey T, Maggs AC, Leibler S 1997. Nature {\bf 389}:305–308.

\bibitem{gonci} G\"{o}nci B, Nagy M, Vicsek T. 2008. Eur. Phys. J. Spec. Top. {\bf 157}:53

\bibitem{aldana} Aldana M {\etal} 2007. Phys Rev Lett {\bf 98}:095702 

\bibitem{chateetal2007} Chat\'e H, Ginelli F and Gr\'egoire 2007. Phys Rev Lett {\bf 99}:229601

\bibitem{kulinskii} Kulinskii VL and Chepizhko AA 2009.  http://arxiv.org/abs/0910.5707

\bibitem{vicsekcsirokbarabasi} Czir\'ok A {\etal} 1999. Phys. Rev. Lett. {\bf 82}:209

\bibitem{raymond_evans} Raymond JR and Evans MR 2006. Phys Rev E {\bf 73}:036112  

\bibitem{couzin1d} Buhl J {\etal} 2006. Science {\bf 312}:1402

\bibitem{vicsekkerato} Szab\'{o} B {\etal} 2006. Phys. Rev. E {\bf 74}:061908

\bibitem{kudpre2003}  Blair DL {\etal} 2003. Phys Rev E {\bf 67}:031303 

\bibitem{aransonrmp} Aranson IS and Tsimring LS 2006. Rev. Mod. Phys. {\bf 78}:641

\bibitem{volfson} Volfson D {\etal} 2004. Phys. Rev. E {\bf 70}:051312

\bibitem{dorbolo} Dorbolo S {\etal} 2005.Phys. Rev. Lett. {\bf 95}:044101 

\bibitem{kudrolliprl100} Kudrolli A {\etal} 2008. Phys. Rev. Lett. {\bf 100}:058001

\bibitem{vjnmsrjstat} Narayan V {\etal} 2006. JSTAT P01005 

\bibitem{gruler} Gruler H {\etal} 1999 Eur. Phys. J B {\bf 11}:187192.

\bibitem{curie} Curie P 1894. J. Phys. III (Paris) {\bf 3}:393

\bibitem{shradhaRG} Mishra S, Simha RA, Ramaswamy S 2010. J. Stat. Mech. P02003 


\bibitem{chateginellimontagne} Chat\'e H {\etal} 2006. Phys. Rev. Lett. {\bf 96}:180602 

\bibitem{veerman} Veerman JAC, Frenkel D 1992. Phys. Rev. A {\bf 45}:5632

\bibitem{shradhasr} Mishra S, Ramaswamy S 2006. Phys. Rev. Lett. {\bf 97}:090602 

\bibitem{dasbarma} Das D, Barma M, Majumdar SN 2001. Phys. Rev. E {\bf 64}:046126

\bibitem{fangfu} Tsai JC, Ye F, Rodriguez J, Gollub JP, Lubensky TC 2005. Phys.
Rev. Lett. {\bf 94}:214301 

\bibitem{pedleykessler} Pedley TJ and Kessler JO 1992. Annu. Rev. Fluid Mech. {\bf 24}:313 

\bibitem{pumplong} Manneville JB, Bassereau P, Ramaswamy S, Prost J. 2001. Phys. Rev. E {\bf 64}:021908

\bibitem{voit} Voituriez R {\etal} 2005. Europhys. Lett. {\bf 70}:404

\bibitem{srmrnjp} Ramaswamy S and Rao M 2007. New J Phys {\bf 9}:423

\bibitem{llelast} Like Euler's buckling instability, see Landau L D, Lifshitz E M 1986. Theory of Elasticity (New York: Pergamon)

\bibitem{giomi} Giomi L {\etal} 2008. Phys Rev Lett {\bf 101}:198101 
 
\bibitem{sumithra} Sankararaman S and Ramaswamy S 2009. Phys. Rev. Lett. {\bf 102} 118107

\bibitem{aransonthin} Aranson IS {\etal} 2007. Phys Rev E {\bf 75}:040901(R)

\bibitem{muhuri} Muhuri S {\etal} 2007. Europhys. Lett. {\bf 78}:48002

\bibitem{shelley1} Saintillan D, Shelley MJ: 2007. Phys Rev Lett {\bf 99}:058102  

\bibitem{shelley2} Saintillan D, Shelley MJ: 2008. Phys Rev Lett {\bf 100}:178103  


\bibitem{ishikawapedley} Ishikawa T, Pedley TJ 2008. Phys. Rev. Lett. {\bf 100} 088103 

\bibitem{ishikawareview} Ishikawa T 2009. J. R. Soc. Interface {\bf 6}:815 

\bibitem{locsei} Ishikawa T, Locsei JT, Pedley TJ 2008. J Fluid Mech {\bf 615}:401–431

\bibitem{mehandianott} Mehandia V, Nott PR 2008. J. Fluid Mech. {\bf 595}:239–264.

\bibitem{graham2005prl} Hern\'andez-Ortiz JP {\etal} 2005. Phys Rev Lett {\bf 95}:203501  

\bibitem{ganesh} Subramanian G. and Koch DL 2009. J Fluid Mech {\bf 632}:359  

\bibitem{ marenduetalpre2007} Marenduzzo D, Orlandini E, Cates ME, Yeomans JM 
2007. Phys. Rev. E {\bf 76}:031921

\bibitem{marenduetaljnnfm} Marenduzzo D, Orlandini E, Cates ME, Yeomans JM 
2008. J. Non-Newtonian Fluid Mech. {\bf 149}:56-62

\bibitem{llopis} Llopis I, Pagonabarraga I 2006. Europhys. Lett. {\bf 75}:999-1005

\bibitem{sanoop} Ramachandran S Kumar PBS, Pagonabarraga I. 2006. Eur. Phys. J. E {\bf 20}:151-158

\bibitem{wolg} Wolgemuth C 2008. Biophys. J. {\bf 95}:1564 

\bibitem{marendu} Marenduzzo D, Orlandini E, Yeomans JM. 2007. Phys. Rev. Lett. {\bf 98}:118102

\bibitem{activerheo} Hatwalne YV, Ramaswamy S, Rao M, Simha RA 2004. 
Phys. Rev. Lett. {\bf 92}:118101

\bibitem{tanniecrisactiverheo} Liverpool TB, Marchetti MC 2006. Phys. Rev. Lett. {\bf 97}:268101 

\bibitem{catesfielding} Cates ME,  Fielding SM, Marenduzzo D, Orlandini E, Yeomans JM. 2008. Phys. Rev. Lett. {\bf 101}:068102 

\bibitem{saintillanexpmech} Saintillan D 2009. Experimental Mechanics DOI 10.1007/s11340-009-9267-0 

\bibitem{prostbruinsma1} Prost J, Bruinsma R  1996. Europhys. Lett., {\bf 33}:321-326

\bibitem{pumppramana} Ramaswamy S, Toner J, Prost J 1999. Pramana {\bf 53}:237

\bibitem{pumpprl} Ramaswamy S, Toner J, Prost J 2000.  Phys. Rev. Lett. {\bf 84}:3494-3497

\bibitem{cras} Ramaswamy S, Rao M 2001. C.R. Acad. Sci. Paris {\bf t. 2, S\'er. IV}:817-839

\bibitem{gautammembrane} Sankararaman S,Menon GI, Kumar PBS 2002. 
Phys. Rev. E. {\bf 66}:031914

\bibitem{xlwu} Wu XL, Libchaber A 2000. Phys. Rev. Lett. {\bf 84}:3017-3020

\bibitem{lau1} Chen DT {\etal} 2007. Phys. Rev. Lett. {\bf 99}:148302 

\bibitem{lau2} Lau AWC 2009 and Lubensky TC 2009. Phys. Rev. E {\bf 80}:011917 

\bibitem{grahamprl2008} Underhill PT, Hernandez-Ortiz JP, Graham MD 2008. Phys. Rev. Lett. {\bf 100}:248101

\bibitem{raminactive} Golestanian R 2009. Phys. Rev. Lett. {\bf 102}:188305 

\bibitem{abhiketal} Basu A, Joanny JF, J\"ulicher F, Prost J 2008. Eur. Phys. J E {\bf 27}:149-160. 

\bibitem{lighthillfish} Lighthill MJ 1975. Mathematical Biofluiddynamics (SIAM, Philadelphia)

\bibitem{weihs} Weihs D 1973. Nature {\bf 241}:290 

\bibitem{makris} Makris NC, Ratilal P, Symonds DT, Jagannathan S, Lee S, Nero RW 2006. 
Science {\bf 311}:660 

\bibitem{kruseetalnjp} Joanny JF, J\"ulicher F, Kruse K, Prost J 2007. New J Phys {\bf 9}:422 

\bibitem{hfsp} Joanny JF, Prost J 2009. HFSP Journal {\bf 3}:94 

\bibitem{takiguchi} Tanaka-Takiguchi Y {\etal} 2004. J. Mol. Biol. {\bf 341}:467 

\bibitem{degroot} De Groot SR, Mazur P 1984. Non-Equilibrium Thermodynamics
(Dover Publications, New York)

\bibitem{salbreux} Salbreux G {\etal} 2009. Phys Rev Lett {\bf 103}:058102 

\bibitem{pramodosc} Salbreux G, Joanny JF, Prost J, Pullarkat PA 2007. Physical Biology {\bf 4}:268.

\bibitem{fdt} Kubo R {\etal} 1991. Statistical Physics II, Nonequilibrium
Statistical Mechanics (Springer, Berlin) 

\bibitem{mizuno} Mizuno D {\etal} 2007. Science {\bf 315}:370-373.

\bibitem{alexfredegennesissue} Levine AJ, MacKintosh FC 2009. J. Phys. Chem. B {\bf 113}:3820 

\bibitem{noriopap} Kikuchi N., Ehrlicher A, Koch D, K\"{a}s JA, Ramaswamy S, Rao M.  2009. Proc Natl Acad Sci USA {\bf 106}:19776-19779 

\bibitem{hudspeth} Martin P, Hudspeth AJ, J\"{u}licher F 2001. Proc. Natl. Acad. Sci. USA {\bf
98}:14380-14385

\bibitem{pramod} Bernal R, Pullarkat PA, Melo F 2007. Phys. Rev. Lett. {\bf 99}:018301

\bibitem{pramodphysrep} Pullarkat PA {\etal} 2007. Phys Rep {\bf 449}:29

\bibitem{jitumadan} Goswami D {\etal} 2008. Cell {\bf 135}:1085-1097 

\bibitem{krusejulicherprl2000} Kruse K, J\"ulicher F 2000. Phys. Rev. Lett. {\bf 85}:1778-1781  
  
\bibitem{cris1} Liverpool TB, Marchetti MC 2003. Phys Rev Lett {\bf 90}:138102.

\bibitem{aparnacristinaPNAS} Baskaran A and Marchetti MC 2009. PNAS 106:15567 

\bibitem{rafai} {Rafa\"\i} S, Peyla P, Jibuti L 2010. Phys. Rev. Lett. {\bf 104}:098102

\bibitem{srcmjc} Ramaswamy S 2009. {\tt http://www.condmatjournalclub.org/?p=760}

\bibitem{sok} Sokolov A, Aranson IS 2009. Phys Rev Lett {\bf 103}:148101

\bibitem{dombrowski} Dombrowski C, Cisneros L, Chatkaew S, Goldstein RE,Kessler JO 2004. 
Phys. Rev. Lett. {\bf 93} 098103

\bibitem{laugareview} Lauga E and Powers T 2009. Rep. Prog. Phys. {\bf 72}:096601 

\bibitem{sok1} Sokolov A, Aranson IS, Kessler JO, Goldstein RE 2007. Phys. Rev.
Lett. {\bf 98}:158102

\bibitem{cisnerosexpfluids} Cisneros LH {\etal}, Cortez R, Dombrowski C, Goldstein RE, Kessler JO 2007. Exp Fluids {\bf 43}:737-753  

\bibitem{activecolloid2} Kline TR {\etal} 2005. JACS {\bf 127}:17150 

\bibitem{raminetalprl2005} Golestanian R {\etal} 2005. Phys Rev Lett {\bf 94}:220801 

\bibitem{ruecknerkapral} R\"{u}ckner G, Kapral R 2007. Phys Rev Lett {\bf 98}:150603 

\bibitem{lammertetal} Lammert PL, Prost J, Bruinsma R 1996. J. Theor. Biol. {\bf 178}:387-391

\bibitem{dreyfus} Dreyfus R {\etal} 2005. Nature {\bf 437}:862 

\bibitem{snigdha} Thakur S {\etal} 2006. Phys. Rev. Lett. {\bf 97}:115701 

\bibitem{yoshi} Chen YJ {\etal} 2009. Phys. Rev. E {\bf 80}:016303 

\bibitem{vk} Kumar KV, Ramaswamy S, Rao M 2008.  Phys Rev E {\bf 77}:020102(R) 

\bibitem{adrian} Baule A, Kumar KV, Ramaswamy S 2008. JSTAT  P11008. 

\bibitem{bartolo_lauga_shaken} Bartolo D and Lauga E 2009. Phys. Rev E 2010. {\bf 81}:026312

\bibitem{starflag} Ballerini M {\etal} 2008. Animal Behaviour {\bf 76}:201-215; http://arxiv.org/abs/0802.1667

\bibitem{jfjptissue} Basan M, Risler T, Joanny J-F, Sastre-Garau X, Prost J 2009. HFSP J. 
{\bf 3}:265-7

\bibitem{boristissue} Shraiman BI 2005. Proc. Natl. Acad. Sci. U.S.A. {\bf 102}:3318-23

\bibitem{durotaxis} Lo CM {\etal} 2000. Biophys. J. {\bf 79}:144

\bibitem{safranmechanosens} Nicolas A {\etal} 2004. Proc. Natl. Acad. Sci. USA
{\bf 101}:12520-12525 

\bibitem{sandlizard} Maladen R {\etal} 2009. Science {\bf
325}:314 
 
\bibitem{laugaviscoel} Lauga E 2007. Phys. Fluids {\bf 19}:083104

\bibitem{bioconvection} Hill NA, Pedley TJ 2005. Fluid Dyn Res {\bf 37}:120

\bibitem{sanoPLoS1} Maeda YT, Inose J, Matsuo MY, Iwaya S, Sano M 2008. PLoS One 3:e3734

\end{thebibliography}
\end{document}